\documentclass[%
 aip,
 amsmath,amssymb,
preprint,%
]{revtex4-1}

\usepackage{graphicx}
\usepackage{dcolumn}
\usepackage{bm}

\usepackage[utf8]{inputenc}
\usepackage[T1]{fontenc}
\usepackage{mathptmx}
\usepackage{etoolbox}
\usepackage{soul,xcolor}

\makeatletter
\def\@email#1#2{%
 \endgroup
 \patchcmd{\titleblock@produce}
  {\frontmatter@RRAPformat}
  {\frontmatter@RRAPformat{\produce@RRAP{*#1\href{mailto:#2}{#2}}}\frontmatter@RRAPformat}
  {}{}
}%
\makeatother

\usepackage{lineno}
\graphicspath{{figures/}}
\usepackage{epstopdf}
\AppendGraphicsExtensions{.tif}
\begin{document}

\preprint{AIP/123-QED}

\title{Partially Saturated Granular Flow in a Rotating Drum: The Role of Cohesion}

\author{Mingrui Dong}
\affiliation{School of Civil Engineering, The University of Sydney, NSW 2006, Australia}
\author{Zhongzheng Wang}%
\affiliation{School of Mechanical, Medical and Process Engineering, Faculty of Engineering, Queensland University of Technology, QLD 4001, Australia
}%
\author{Benjy Marks}
\affiliation{School of Civil Engineering, The University of Sydney, NSW 2006, Australia}
\author{Yu Chen}
\affiliation{School of Civil Engineering, The University of Sydney, NSW 2006, Australia}

\author{Yixiang Gan}
\affiliation{School of Civil Engineering, The University of Sydney, NSW 2006, Australia}
\affiliation{Sydney Nano, The University of Sydney, NSW 2006, Australia}
\email{yixiang.gan@sydney.edu.au}

\date{\today}

\begin{abstract}
\section*{Abstract}
Partially saturated granular flows are common in various natural and industrial processes, such as landslides, mineral handling, and food processing. We conduct experiments and apply the Discrete Element Method (DEM) to study granular flows in rotating drums under partially saturated conditions. We focus on varying the strength of cohesion (surface tension) and rotation rate within the modes of rolling flow and cascading flow. With an increase in surface tension, a rolling mode can possess a steeper slope and correspondingly needs a higher rotation rate to transition to a cascading. The depth of the flowing region increases with increasing cohesion, while the sensitivity is reduced for cases of high cohesion.
We propose a dimensionless number $C_E$ that captures the combined effects of rotation, gravity and cohesion on the dynamic angle of repose and flow depth.
In addition, we extract statistical information on the formation of clusters within the flow. We find a power law relation between the cluster size distribution and its probability, which indicates that stronger cohesion can promote the formation of larger clusters, and we discuss how cohesion impact on flows manifested by cluster formation.
%
\end{abstract}

\maketitle

\section*{Highlights}
\begin{itemize}
\item Partially saturated granular flow is investigated using rotating drum experiments and the Discrete Element Method (DEM).
\item A dimensionless number ($C_E$) is proposed based on inertia, gravity, and cohesion to characterise the cohesive granular flow.
\item A transition from rolling to cascading flow is observed due to changes in cohesion and can be characterised by the morphology and depth of the flowing region.
\item The formation of clusters due to cohesion is analysed and its impacts on the depth of the flow is highlighted.
\end{itemize}

\newpage

\section{\label{sec:intro}Introduction}

Granular materials are prevalent in all domains of human activity, from natural sands to industrial and mining operations \cite{andreotti2013granular}. When disturbed, these materials can transition from behaving as a solid to flowing like a liquid. The rheological properties of this flowing material have been studied under various conditions \cite{komatsu2001creep,gdr2004dense,forterre2008flows,norouzi2015insights,espiritu2020investigation}. Individual grains in these flows can be characterised by properties such as their size, shape and mass. A granular flow, when described as a continuum, can alternatively be described by fields such as the density (or solid fraction), pressure and strain rate. To determine the relationships between the individual particle properties and their continuum equivalents, many apparatus and simulations with different geometries have been studied for granular media with and without the presence of inter-particle cohesion \cite{gdr2004dense,rognon2008dense,forterre2008flows,orpe2001scaling,komatsu2001creep,wang2019granular,mandal2020insights,bonamy2002multiscale,orpe2007rheology,forterre2008flows,cortet2009relevance,jarray2019wet}. However, there is still a lack of comprehensive understanding of the flow behaviour of cohesive granular materials.

A granular surface flow can occur due to the presence of gravity and a difference in height. A surface flow is characterised by a flowing layer of grains at the surface, with stationary grains underneath. Surface particles flowing along steeper slopes tend to possess higher velocities \cite{pouliquen2004velocity}. For slightly cohesive particles under the same conditions, surface particle velocity is observed to be reduced \cite{jarray2019wet}. With the increase of cohesion, the surface flow can transition to a plug flow where a plug region can be found overlying a shear band  \cite{brewster2005plug,rognon2008dense,mandal2020insights}. The stratification of the flow region can be viewed as a transition from a collisional flow to a viscoplastic regime \cite{da2005rheophysics,mandal2020insights,tegzes2002avalanche}. The variation of the flowing layer depth, \textit{i.e.}, the distance between the flow surface and the flow-solid interface, may indicate the dissipation of the amount of gravitational energy that was previously stored by particles flowing on the top \cite{chou2011experimental, jarray2019wet}.

Since characteristic particle size plays a crucial role in affecting various aspects of granular flow, including inter-particle friction, packing and porosity, permeability and seepage, segregation and mixing \cite{renouf2005numerical,ostoja2006material,qu2021adaptive,shaheen2021influence}, a description of the cluster formation can be critical\cite{bonamy2002multiscale,pouliquen2004velocity,fullmer2017clustering}. It is particularly useful to understand the flow mechanism of cohesive particles\cite{tegzes2002avalanche,gdr2004dense,pouliquen2004velocity,rognon2008dense,cao2018structural}. \citet{bonamy2002experimental} studied cluster formation in cohesionless particle surface flows based on velocity fluctuations, where a negative power law relation was identified between the cluster size (quantified by the number of particles within a cluster) and their corresponding probabilities. \citet{cao2018structural} demonstrated that the inter-cluster plastic rearrangement between unstable tetrahedral clusters plays an essential role in the plasticity of sheared granular materials. \citet{zou2022microscopic} presented how the particle shape (aspect ratio) can contribute to increasing the cluster formation in a dense granular flow. For cohesive particles, a viscoplastic flow accompanied by free surfaces that possess repetitive patterns due to the formed clusters can be observed \cite{tegzes2002avalanche}. The velocity of intra-cluster particles remains coherent, which can lead to a relatively low local granular temperature \cite{jarray2019wet}. \citet{rognon2008dense} mentioned that clusters can play a key role in the dilation effect of cohesive granular flows.

The rotating drum has been widely used in various applications and provides rich information on granular flow \cite{bonamy2002multiscale,orpe2007rheology,forterre2008flows,cortet2009relevance,norouzi2015insights,jarray2019wet}. Particles in a rotating drum can be divided into two parts. Firstly, a passive, "solid-like", region is rotated along the tube before flowing down. Secondly, an active flow, "fluid-like", region forms a surface flow. As the rotation rate increases for a given granular medium, the flow can be characterised into six surface flow modes; slipping (no flow), slumping (intermittent avalanches), rolling (continuous flow with a tilted flat surface), cascading (continuous flow with a curved "$S$" shape surface), cataracting (overturning tumbling) and centrifuging (all particles stuck to the tube walls) \cite{henein1983experimental}. It has been shown that the transition between these modes depends on a variety of parameters, \textit{e.g.}, the rotation rate, particle shape, particle-to-drum size ratio, filling rate, wall effects, and cohesion effects \cite{nowak2005maximum,taberlet2006s,arntz2008granular,brewster2009effects,liu2013self,jarray2019wet}. The dimensionless Froude number ($Fr$) is widely adopted to describe the mode transition,
\begin{equation}\label{eq:Fr}
    Fr=\sqrt{\omega^2 R/g},
\end{equation}
where $\omega$ is the rotation rate, $R$ the drum radius, and $g$ the gravitational acceleration. The Froude number considers the competing effects of inertia and gravity \cite{arntz2008granular,jarray2019wet}. A low Froude number typically corresponds to slipping, and a Froude number greater than one represents centrifuging. In this work, unless stated otherwise, we focus on flow transition between rolling and cascading flow modes. Within the rolling and cascading modes, the dynamic angle of repose $\theta$ is a critical parameter for characterising the behaviour of the granular flow \cite{henein1983experimental,xu2007lubrication,jarray2019wet,chen2021elastic}. A relationship between $Fr$ and the dynamic angle of repose can be established\cite{rajchenbach1990flow,arntz2008granular} where an increase of $Fr$ can drive the transition gradually from slumping to rolling and then to cascading.

For cohesive particles in a rotating drum, two additional dimensionless numbers become relevant, namely, the Bond number ($Bo$), which describes the relative importance of cohesion compared to the gravitational force, and the Weber number ($We$), representing the ratio of inertia to cohesion\cite{nowak2005maximum,jarray2019wet}. These two dimensionless numbers are expressed as:
\begin{align}
    Bo &= \frac{\gamma \cos{\theta_C}}{\rho g r^2_p},\label{eq:Bo}\\
    We &= \frac{\rho r_p (\omega R)^2}{\gamma\cos{\theta_C}},\label{eq:We}
\end{align}
where $\gamma$ is the liquid surface tension, $\theta_C$ the multiphase contact angle, $\rho$ the material density, $r_p$ the average particle radius, and $\omega$ the rotation rate. Within relatively high cohesion, typically, the $Bo$ > 1, intermittent avalanches can be observed when the rotating rate is low \cite{elekes2021expression}. Whilst under the same condition, a continuous flow can be found for cohesionless particles\cite{nowak2005maximum}. Within continuous flow modes, a nonlinear correlation can be found between $We$ and the dynamic angle of repose \cite{jarray2019wet}.

Here, we investigate the behaviour of partially saturated granular media in a rotating drum through experimentally validated simulations with the Discrete Element Method. We focus on quantifying and analysing three key characteristic features; the dynamic angle of repose ($\theta$), the flow depth ($h$), and the formation of clusters, under various rotation rates ($\omega$) and surface tensions ($\gamma$). After the introduction, we first present the experimental setup and numerical method in Section \ref{sec:method}. In Section \ref{sec:results}, we discuss the effect of cohesion on these characteristic features to understand the relationship between cohesion energy (provided by $\gamma$) and kinetic energy (by $\omega$). We propose a dimensionless number that takes gravity, inertia, and cohesion into account, and demonstrate its effectiveness in characterising cohesive granular flows in a rotating drum. Finally, we discuss the implications of our findings and conclude with potential research directions. 

\section{\label{sec:method}Method}

\subsection{Rotating drum setup}
\begin{figure*}
\includegraphics[width=0.8\textwidth,height=\textheight,keepaspectratio]{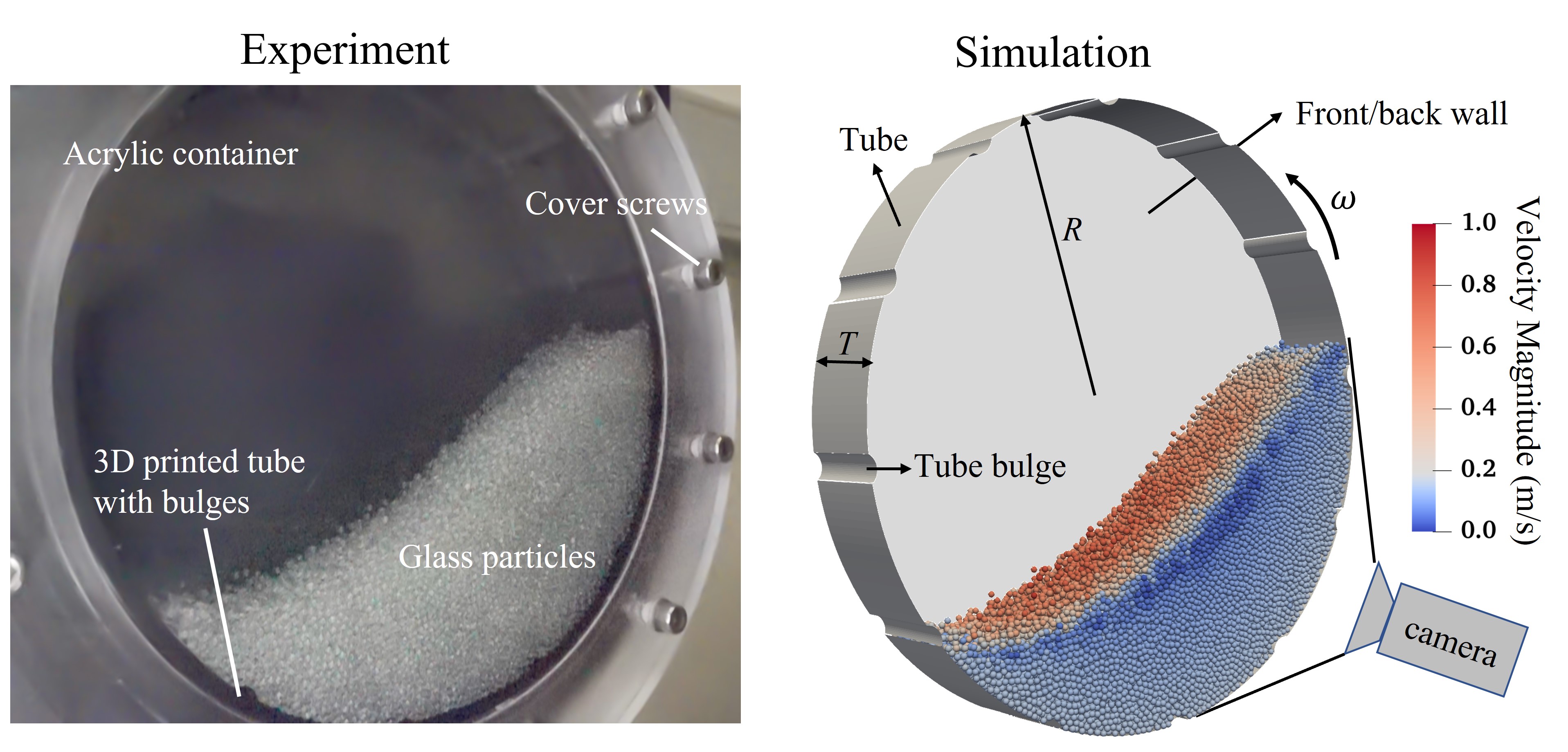}
\caption{The experiment and simulation setup. \emph{Left}: A photo of the experimental rotating drum with a rotation rate $\omega = 15$~rpm under dry conditions. \emph{Right}: A simulation of the same scenario as the left experimental photo. The rotating drum has a radius $R$ and depth $T$. Twelve bulges are used to avoid slipping between particles and the drum. The drum rotates anti-clockwise. The colour scale from blue to red represents an increase in particle velocity Magnitude.}
\label{fig:simulation_setup}
\end{figure*}

A photograph of the experimental apparatus is shown in Fig.~\ref{fig:simulation_setup}. The sidewalls of the rotating drum are fabricated from clear acrylic. The drum body is 3D printed polylactic acid thermoplastic (density 1.24~g/cm$^3$) using an Ultimaker3 3D printer. Twelve bulges are spaced evenly around the drum to prevent slip between particles and the drum. The bulges are cylindrical with diameter $2r_p$ which ensures that their effect on the flow behaviour in bulk regions is negligible \cite{sunkara2013influence,zhang2018experimental}. The system is made watertight with a sealing strip between the drum and the sidewalls. Two non-stick and transparent FEP fluoropolymer films (with low surface energy leading to a contact angle of around $100^\circ$) are attached to the sidewalls to prevent wet particles from sticking to them. Glass beads with average radius $r_p \approx 1.01~\mathrm{mm}\pm3\%$ are used. A high-speed camera shooting at a frame rate of $500~\mathrm{fps}$ with 1920$\times$1080 pixels is used to capture the granular flow through the transparent sidewalls. A LED ring light directed at the drum is set to provide enough light for the camera, and a blackboard is attached to the back of the drum to act as a homogeneous contrasting background. At this zoom level, each particle is approximately $8\times8$ pixels, which ensures precision of the experimental results.

The filling level of the drum is expressed as $\tfrac{V_\mathrm{filled}}{V_\mathrm{container}}\approx30\%$, where a total of 376.6~g of glass beads in the experiment is used. We study here low volumetric water content conditions, \textit{i.e.} $m_w=V_\mathrm{liquid}/V_\mathrm{solid}~\leq~2\%$, where $V_\mathrm{liquid}$ is the volume of liquid and $V_\mathrm{solid}$ is the volume of solids. In this state, the liquid exists mainly in the form of liquid bridges in the granular materials, which creates attractive inter-particle (capillary) forces and consequently impacts the behaviour of the granular medium \cite{feng1998effect,scheel2008morphological,wang2021packing}. In this study, we consider the capillary force supplied by the liquid bridge as the primary source of this cohesion effect, which is the case for particles of size greater than \textasciitilde100~µm where van der Waals force can be neglected \cite{mutsers1977effect,feng1998effect}. In the experiment, 3.1~mL of distilled water is added to make \textasciitilde2\% volumetric water content in the experiment to ensure a pendular state.

A rendering of the simulation is shown in Fig.~\ref{fig:simulation_setup}. We use a drum with internal radius $R = 96r_p$ and thickness $T = 30r_p$. 37,500 particles are used in the simulation to ensure the filling level is around $30\%$. The detail of the DEM modelling is explained in Subsection~\ref{subsec:contactlaw}.

\subsection{Discrete Element Method (DEM)}\label{subsec:contactlaw}
The Discrete Element Method (DEM) has been proven to be effective in investigating granular phenomena with abundant detail \cite{cundall1979discrete,walton1993numerical,luding2005anisotropy,gan2010discrete,gdr2004dense}. Here, we use the open-source platform LIGGGHTS to simulate the granular system~\cite{kloss2012models}. The particle movement in DEM is described by the computation of Newton's second law. The normal, tangential, and rolling inter-particle contact laws are respectively described as \cite{brilliantov1996model,iwashita1998rolling,ai2011assessment}:
\begin{align}\label{eq:1}
F_n^{\mathrm{Hertz}} &= k_n\delta_n - \beta_n\nu_n,\\
F_t &= k_t\delta_t - \beta_t\nu_t,\\
M_r &= k_r\Delta\theta_r - \beta_r\omega,
\end{align}
where $n$, $t$, and $r$ represent the normal, tangential, and rolling directions, the superscript "Hertz" represents the Hertzian contact, $F$ the contact force, $k$ the contact stiffness, $M$ the rolling torque, $\delta$ the overlap distance between contacting particles, $\Delta\theta_r$ the relative rotation angle of contacting particles, $\beta$ the coefficient of damping, and $\nu$ and $\omega$, respectively, denote the relative translational velocity and relative angular velocity for contacting neighbours. Here, $k_n$ is expressed as the nonlinear Heartzian law by $k_n=\frac{4}{3}E\sqrt{R^*\delta_n}$, where $E$ is the Young's modulus, $R^*=\frac{r_ir_j}{r_i+r_j}$ the effective radius of contacting neighbours. The tangential friction force and rolling torque, $F_t$ and $M_r$, are capped to satisfy a Coulomb style criteria defined by $|F_t|\leq\mu_s|F_n^{\mathrm{total}}|$ and $|M_r|\leq\mu_r r_pF_n^{\mathrm{total}}$ where $\mu_s$ is the coefficient of sliding friction and $\mu_r$ is the coefficient of rolling friction \cite{iwashita1998rolling}. Notice that $\mu_s$ for glass beads can vary from 0.2 to 0.65 \cite{parteli2014attractive,yang2000computer,fuchs2014rolling,tang2019measurement,bhateja2016scaling}, we select $\mu_s = 0.5$ in this study, a common value adopted in the literature \cite{parteli2014attractive,yang2000computer}. A value of $e = 0.9$ is selected as the coefficient of restitution which is the same as the previous study by \citet{bhateja2016scaling}. Computational stability \cite{gan2010discrete,cundall1979discrete} is guaranteed by checking the time step $t = 10^{-8}s<\sqrt{m/Ed_s}$, where $m$ denotes the mass of the smallest particle, $d_s=0.95d_p$ denotes the diameter of the smallest particles, and particle size in this study is set uniformly distributed as $d_p=$ 2 $\pm 5 \%$. Table \ref{tab:Model parameters} lists key parameters adopted in this study.

The constitutive model of capillary force proposed by \citet{soulie2006influence} is described by
\begin{equation} \label{eq:2}
    F_n^{\mathrm{cap}} = \pi\gamma\sqrt{r_i r_j}[c + e^{(a\delta_\mathrm{gap}/r_j + b)}],
\end{equation}
where $F_n^{\mathrm{cap}}$ is the normal capillary force, $r_i$ and $r_j$ are the radius of two neighbouring particles linked by a liquid bridge with $j$ indicating the larger one, $\delta_\mathrm{gap}$ is the distance between the surfaces of each pair of neighbouring particles, $a$, $b$ and $c$ are coefficients deduced from the liquid bridge volume ($V_{ij}^{\mathrm{bri}}$), $\theta_C$ and $R_j$ \cite{soulie2006influence}. Here, the liquid film is assumed\cite{willett2000capillary} to fully cover the surface of particles, which leads to $\theta_C = 0^\circ$. A liquid bridge can form when the surface gap distance reaches $D_{\mathrm{birth}} = h_{t,i} + h_{t,j}$, which takes the liquid film thickness $h_{t,i}=\sqrt[3]{V_i^\mathrm{solid}+V_i^\mathrm{liquid}} - r_i = 0.007r_i$ into consideration. The liquid bridge rupture distance $D_{\mathrm{rupture}} = (1 + 0.5\theta_C)\times(V_{ij}^{\mathrm{bri}})^\frac{1}{3}$ is adopted, beyond which the liquid bridge breaks and $F_n^{\mathrm{cap}}$ no longer exists \cite{lian1993theoretical}. Based on the assumption $m_w\le2 \%$, the liquid volume on each particle is set as $V_i^\mathrm{liquid} = V_i\times2\%$, where $V_i$ denotes the volume of the particle $i$. The liquid bridge volume $V_i^{\mathrm{bri}}$ is calculated as $0.05\times(V_i^\mathrm{liquid} + V_j^\mathrm{liquid}) = (V_i + V_j)\times0.1\%$, where the coefficient 0.05 ensures liquid volume conservation when contacting neighbours of each particle may reach twelve (face-centred-cubic topological structure), meaning a maximum of $60\%$ of the effective volumetric liquid content of one particle ($V_i^\mathrm{liquid}$) forms liquid bridges. Note, the liquid weight plays a negligible role in the dynamics of particles since the mass ratio between liquid and particles is $\rho_l\times\frac{0.02}{\rho_p}\approx2\%$. The particle-wall contact force is assumed to be the same as the particle-particle contact, with the wall side possessing an infinite radius to mimic a plate \cite{taberlet2006s,jaggannagari2021simulations}. A detailed illustration and analysis of the formation and rupture of the liquid bridges under different surface tensions can be found in our previous work\cite{dong2022wet}.

\begin{table}
\caption{\label{tab:Model parameters}Model parameters }
\begin{ruledtabular}
\begin{tabular}{ll}
\textbf{Parameter}   & \textbf{Value} \\ \hline
     Young's modulus, $E$ (Pa)  
     & 6$\times10^{10}$  \\ 
     Coefficient of sliding friction, $\mu_s$ (-)  
     & 0.5  \\ 
     Coefficient of rolling friction, $\mu_r$ (-)  
     & 0.001  \\ 
     Coefficient of restitution (-)  
     & 0.9  \\ 
     Surface tension, $\gamma$ (N/m)   
     & 0  - 0.146 \\ 
     Contact angle, $\theta_C$ ($^\circ$)
     & 0 \\ 
     Gravitational acceleration, $g$ (m/s$^2$)
     & 9.81\\ 
     Liquid dynamic viscosity, $\mu_v$ (Pa$\cdot$s)
     & $8.9\times10^{-4}$\\ 
     Particle diameter, $d_p$ (mm)    
     & 2 $\pm 5 \%$  \\ 
     Particle density, $\rho$ (kg/m$^3$)    
     & 2460 \\ 
     Drum radius / depth, R / T (-)
     & 48$d_p$ / 15$d_p$ \\ 
     Volumetric liquid content, $m_w$ (-)
     & $2\%$ \\
\end{tabular}
\end{ruledtabular}
\end{table}

\section{Results and discussion}\label{sec:results}
In this section, we first present the result of macro-scale phenomena, \textit{i.e.}, mode transition and dynamic angle of repose, in a parametric study by varying $\omega$ and $\gamma$. Through this process, the simulation is validated. After that, characterisation indices of the meso-scale phenomena (flow depth) and the micro-scale phenomena (the size distribution of cohesion-induced clusters) are extracted. We propose a new dimensionless number to gain insight from these characteristic indices to discuss how cohesion affects the granular flow regime in the rotating drum.

\subsection{Flow mode and dynamic angle of repose}\label{subsec:anglerepose}
\begin{figure*}
\centering
\includegraphics[width=1\textwidth,height=\textheight,keepaspectratio]{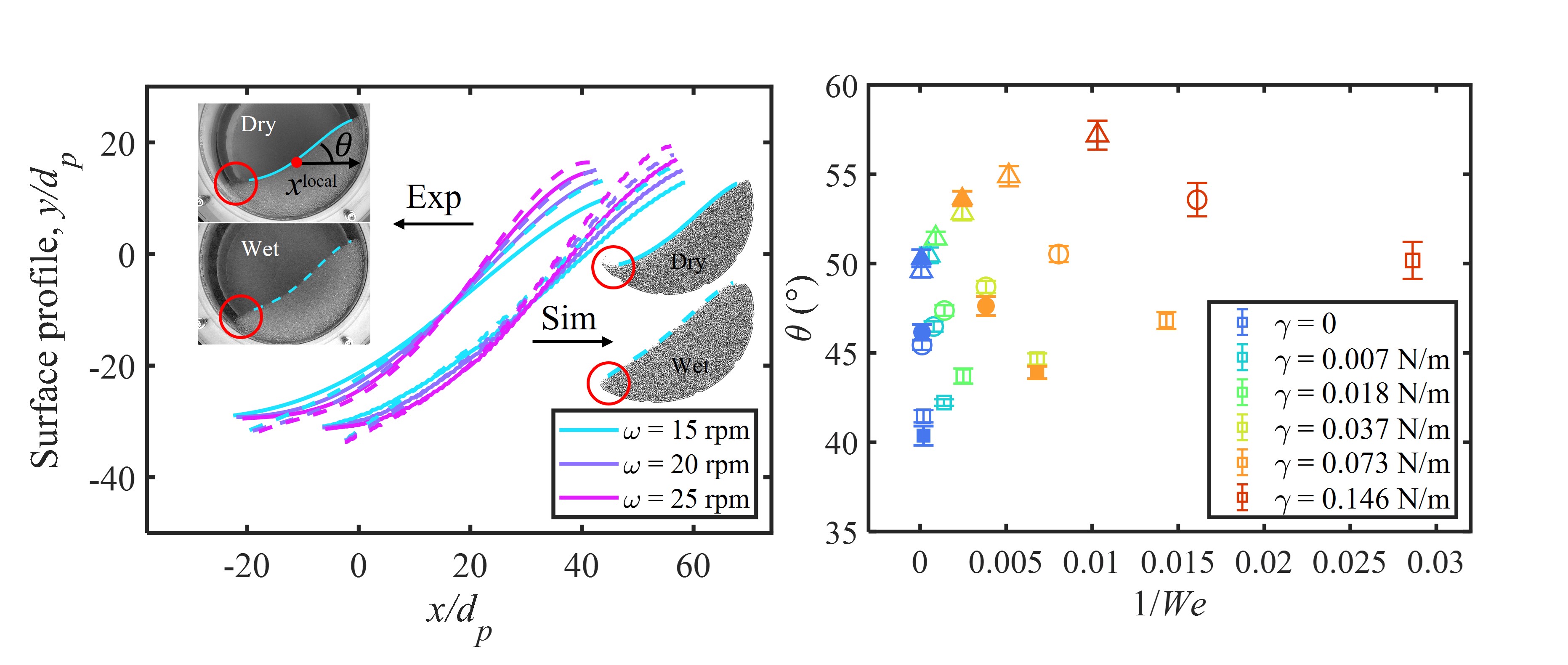}
\caption{Surface profile comparison between experiments and simulations under dry and wet conditions. (a) Surface profiles of experiments (left lines) and simulations (right lines). Solid lines represent dry conditions, and dashed lines wet (cohesive) conditions. Insets at the left show two photographs of the experiment under dry and wet conditions. As a comparison, the right insets show two renderings of simulations under the corresponding conditions to the left insets. Note a slight difference in the filling level (volumetric particle content) between experiments (29.7$\%\pm0.6\%$) and simulations (30.2$\%$) can be the reason that profiles less ideal intersect. Besides, axes centres for both experiments and simulations (Sim) should be identical, however, a shift toward the left is applied to the Sim cases to observe the profile clearly. Red circles highlight the difference in slope toes between cohesive (flattened) and dry (curved) conditions. See the detailed description of how the dynamic angle of repose $\theta$ is defined in Section \ref{subsec:anglerepose}. (b) The dynamic angle of repose $\theta$ as a function of 1/We. Hollow markers are simulation results, and solid markers are experimental results. $\square$, {\Large $\circ$}, and $\triangle$ represent $\omega$ = 15, 20 and 25~rpm respectively.
}
\label{fig:surfaceprofile_DAOR}
\end{figure*}

We investigate the flow regime during the mode transition between rolling and cascading. Note the transition is evaluated through visual observation results \cite{norouzi2015insights,he2019flow}, and the curvature is not quantitatively examined. Both cohesionless (dry) and cohesive (wet) conditions under rotating speed $\omega$ = 15, 20 and 25~rpm are used to validate the simulation against the experiment. As shown in Fig.~\ref{fig:surfaceprofile_DAOR} (a), comparing each of the cohesionless and cohesive cases indicates that the addition of cohesion results in a less curved slope (Cyan dashed lines) with respect to the cohesionless cases (Cyan solid lines). This shows that as $\omega$ increases, the cohesive case presents a delayed transition from rolling to cascading. That is, one needs a relatively higher $\omega$ to enable a rolling-cascading transition for cohesive particles than that for dry particles. An uplifted slope toe of the cohesionless condition can be used to distinguish it from the cohesive case, which is highlighted by red circles in Fig. \ref{fig:surfaceprofile_DAOR} (a). A further increase of cohesion can lead to a flattened profile which was observed in previous studies \cite{liu2011dynamics,jarray2019wet,jaggannagari2021simulations} and will also be presented in Section \ref{subsec:flowdepth}.

\begin{figure}
\centering
\includegraphics[width=0.5\textwidth,height=\textheight,keepaspectratio]{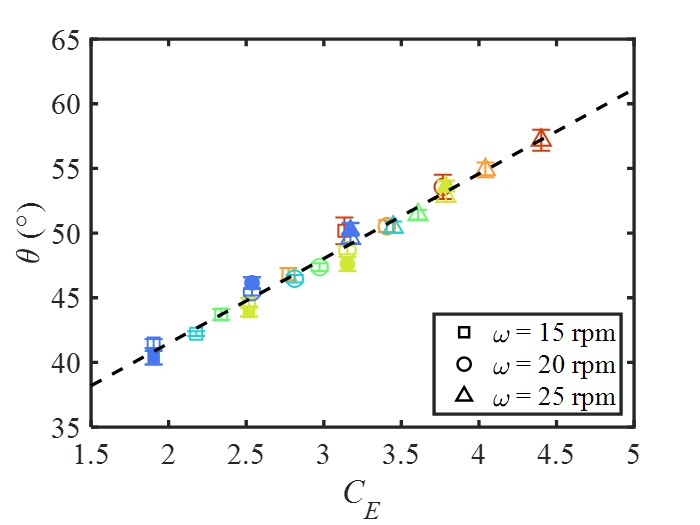}
\caption{Variation of the dynamic angle of repose as a function of the non-dimensional parameter $C_E$. The dashed line represents Eq.~\ref{eq:ce}. Colour transitions from blue to red indicate an increase in cohesion, see Fig.~\ref{fig:surfaceprofile_DAOR} (b).}
\label{fig:DAOR_CE}
\end{figure}

To quantitatively capture the differences in surface profiles under different cohesion and rotation rate, the dynamic angle of repose ($\theta$) is extracted. As illustrated in the top left inset of Fig. \ref{fig:surfaceprofile_DAOR} (a), a coordinate can be constructed taking the red dot as the origin located on the surface with the $x^\mathrm{local}$ axis points towards the right, and an angle between the slope and $x^\mathrm{local}$ can be obtained. When the red dot moves along the slope from the toe to the top, a variation of $\theta$ can be found. Here, Unless stated otherwise, a maximum value of $\theta$, representing the steepest point for the free surface slope, is defined as the dynamic angle of repose \cite{rajchenbach1990flow, jarray2019wet, chen2021elastic}. Note the definition of $\theta$ is the same in both experiments and simulations. The $\theta$ is plotted against the inverse of $We$ in Fig. \ref{fig:surfaceprofile_DAOR} (b), and the $\theta_C\approx60^\circ$ is assumed in the experiment based on previous studies \cite{klise2016automated,wang2021packing}. The relation between $\theta (\omega, \gamma)$ and the inverse of $We$ shows a sharp increase at low cohesion cases followed by a near plateau at high cohesion, which is consistent with previous work \cite{jarray2019wet}. Moreover, in Fig. \ref{fig:surfaceprofile_DAOR} (b), we can see that either an increase in cohesion or rotation rate can lead to a higher $\theta$. 

To measure the impact of the cohesion and the rotation rate on $\theta$, we discuss the influencing factor separately. \citet{nowak2005maximum} examined low rotation rate in the slumping mode with intermittent avalanches, where gravity and cohesion dominate the granular flow. Under these conditions, the maximum static angle of repose could be correlated to the cohesion strength through force balance, leading to the relationship \cite{nowak2005maximum}:
\begin{equation}\label{eq:nowak}
    \mathrm{max}(\theta_\mathrm{stat}) - \mathrm{min}(\theta_\mathrm{stat}) \propto \sqrt{(\frac{9\pi}{2\phi_s})^{\frac{1}{3}}(\frac{\alpha\cos{(\mathrm{min}(\theta_\mathrm{stat}))}\gamma}{\sqrt{6}\tan{(\mathrm{min}(\theta_\mathrm{stat})}\rho gr_pL})}, 
\end{equation}
where $\mathrm{max}(\theta_\mathrm{stat})$ is the maximum static angle of repose which is not affected by the rotating speed, $\mathrm{min}(\theta_\mathrm{stat})=23.8^\circ$ the minimum static angle of repose that can balance particles weight on a slope, $\phi_s$ the solid packing fraction, $\alpha$ a constant coefficient, $r_p$ the particle radius, and $L$ a constant representing a characteristic system size. This equation describes a linear relationship between the maximum static angle of repose with surface tension under a cohesion-dominant condition. Given the definition of Bond number in Eq.~(\ref{eq:Bo}), Eq.~(\ref{eq:nowak}) can be further simplified to
\begin{equation}
    \mathrm{max}(\theta_\mathrm{stat}) - \mathrm{min}(\theta_\mathrm{stat}) \propto Bo^{1/2},
\end{equation}
implying a linear relationship between the increase in the static angle of repose and $Bo^{1/2}$.

At higher relative rotation rates, when inertial effects become dominant, it has been observed that the flow becomes continuous and the $Fr$ can be adopted to evaluate the dynamic angle of repose\cite{jarray2019wet}, which can be expressed as
\begin{equation}\label{eq:jarray}
\theta - \theta_\mathrm{fit} \propto Fr,
\end{equation}
where $\theta_\mathrm{fit} \approx 22^\circ$ is a fitting parameter \cite{jarray2019wet}. Here, Eq.~(\ref{eq:jarray}) also indicates a linear relationship between the increase in the angle of repose and $Fr$.

To attempt to take both the cohesion and inertial effects into account, based on Eqs.~(\ref{eq:nowak}) and (\ref{eq:jarray}), we define a new non-dimensional parameter $C_E=Bo^{1/2} + \lambda Fr$, where $C$ stands for "Combination" and $E$ "Energy Effect", $\lambda$ is a constant independent of $\gamma$, $\omega$, and $g$. Using this definition, we obtain a new expression for the dynamic angle of repose $\theta$ for cohesive granular materials in the granular drum
\begin{equation}\label{eq:ce}
    \theta \propto Bo^{1/2} + \lambda Fr = C_E.
\end{equation}

Fig.~\ref{fig:DAOR_CE} shows a collapse of $\theta$ of all our experimental and numerical results under different rotation rate and surface tensions, demonstrating the validity of the proposed dimensionless parameter $C_E$ in describing the granular flow. The optimal $\lambda$ is fitted in Eq. \ref{eq:ce} using linear regression to be 12 with a goodness of fit $R^2\approx0.97$.

\begin{figure*}
\centering
\includegraphics[width=0.7\textwidth,height=\textheight,keepaspectratio]{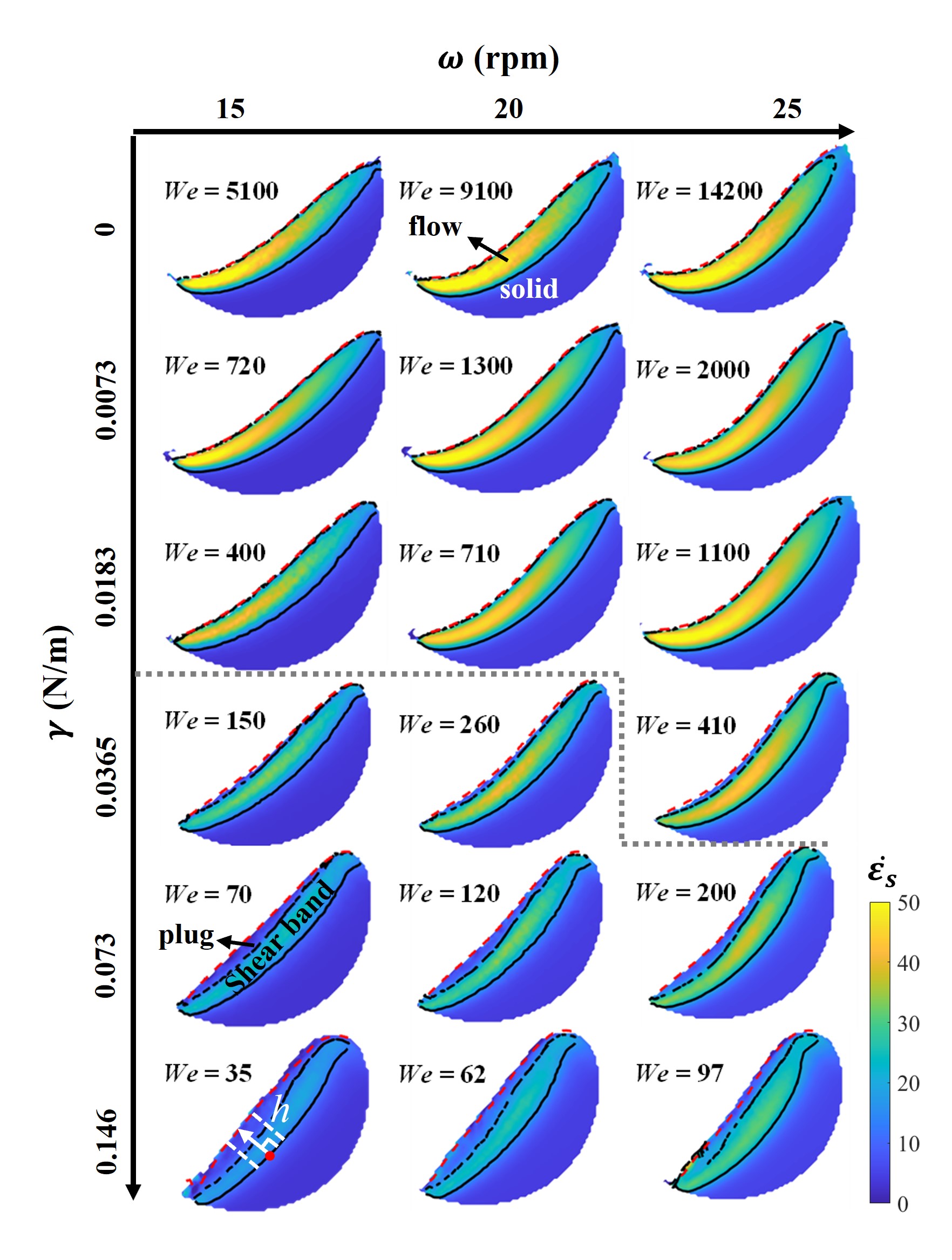}
\caption{Phase diagrams of shear strain rate ($\dot\varepsilon_s$) as a function of $\gamma$ and $\omega$. $We$ is calculated for each case. To obtain $\dot\varepsilon_s$, local particle information, such as particle position and velocity, is obtained using a square sampling mesh of size 2$d_p$. The red dashed line in each inset indicates the measured flow surface. The black solid line is the flow-solid interface, and the black dashed line separates the flow region into a plug zone and a shear zone, both delineated by $\dot\varepsilon_{s,\mathrm{cri}}$. The illustration in the inset $We$ = 35 shows the extraction of the flow depth $h$. Similar to the establishment of the coordinate for $\theta$, here, a coordinate is established with the origin (the red dot) located on the flow-solid interface (the black solid line). As the origin moves along the flow-solid interface, the arrow perpendicular to the interface segment between white dashed lines with length 10$d_p$ can be drawn to obtain the interface-surface distance, which is the flow depth. The grey dotted line indicates the visually observed mode transition from cascading to rolling when $We \approx 400$.}
\label{fig:strrcontour_diagram}
\end{figure*}

\subsection{Flow depth}\label{subsec:flowdepth}
Underneath the surface, the particle assemblies can be divided into flow and solid regions based on the index that depicts the local flow (deformation) with respect to time, \textit{i.e.}, shear strain rate $\dot\varepsilon_s$. Here $\dot\varepsilon_s$ is expressed as $\dot\varepsilon_s=\frac{1}{2}(\frac{\partial v_x}{\partial y} + \frac{\partial v_y}{\partial x})$, where $v_x$ and $v_y$ are time-series averaged local velocities (mapped in mesh grids with a size 2$d_p$) along the $x$ and $y$ axes, and $\partial$ is the differential operator. The shear strain rate for varying cohesion and rotation rate are shown as a phase diagram in Fig. \ref{fig:strrcontour_diagram}. A critical value $\dot\varepsilon_{s, \mathrm{cri}}$ = 10 $\omega$ is applied to indicate the flow-solid interface, which is shown as black solid lines in the rendering of all cases in Fig. \ref{fig:strrcontour_diagram}. 

\begin{figure*}
\centering
\includegraphics[width=1\textwidth,height=\textheight,keepaspectratio]{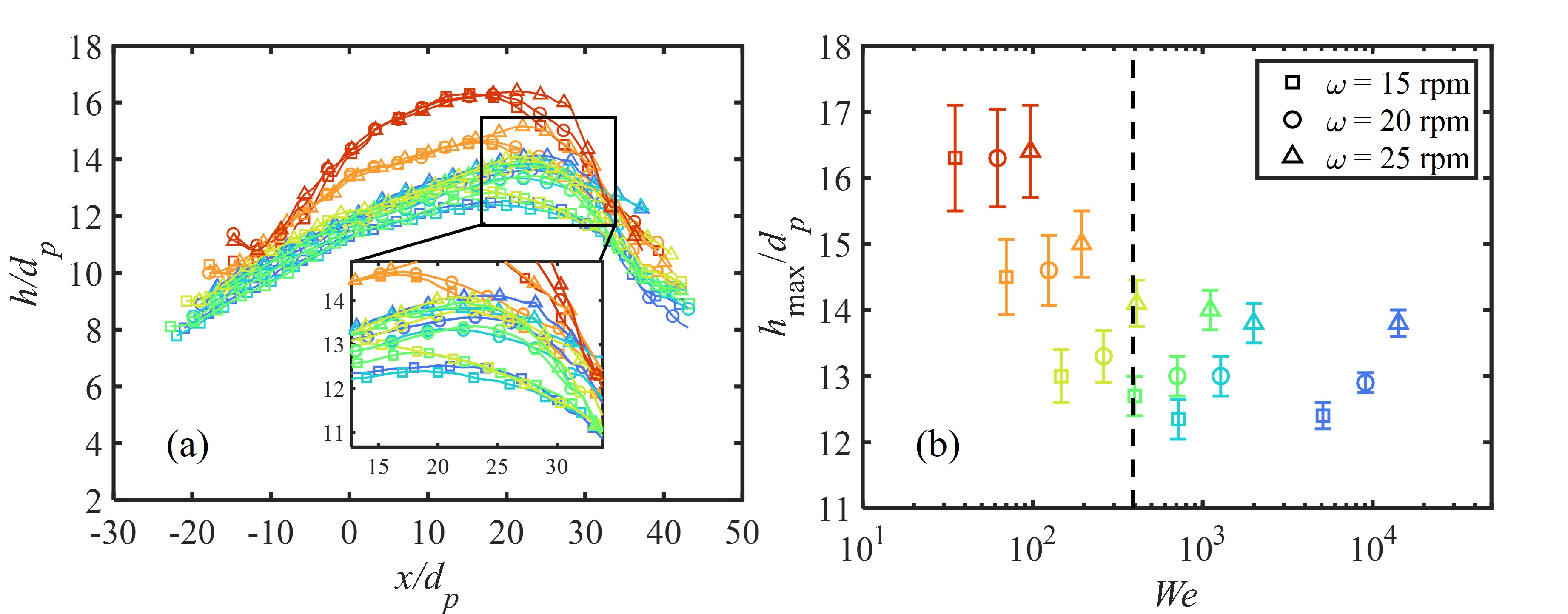}
\caption{(a) The flow depth $h$ scanning along $x$ (horizontal) direction, which is illustrated in the inset ($\omega$ = 15~rpm and $\gamma$ = 0.146~N/m) of Fig. \ref{fig:strrcontour_diagram}. The inset shows the area in the square for clarity. (b) The max flow depth $h_\mathrm{max}$ is plotted as the function of $We$. Colour transitions from blue to red indicate an increase in cohesion, which can be referred to Fig. \ref{fig:surfaceprofile_DAOR} (b). The dashed line shows a critical value of $We$ = 400.}
\label{fig:flowdepth_weber_strainrate}
\end{figure*}

Within the flow region, a relatively high $\dot\varepsilon_s$ area in the lower part of the slope can be observed when cohesion is not dominant. As a comparison, a relatively high cohesion tends to split the flow region into a plug zone overlying a shear band \cite{tegzes2002avalanche,rognon2008dense,mandal2020insights} (Fig. \ref{fig:strrcontour_diagram}, $We$ = 70). The variation of flow region can be attributed to the transition from a collisional regime to a visco-plastic regime due to the increase of cohesion. Within the collisional regime, particles possess relatively high velocities while moving downwards since little energy can be dissipated through collisions. Additionally, the flow region is eroded by the solid region because particles near the flow-solid interface are trapped in the solid region, which narrows the flow region near the toe. The combination of relatively high velocity and narrowed flow region leads to a higher temporal deformation, \textit{i.e.}, high $\dot\varepsilon_s$, near the toe of slopes. In contrast, for the visco-plastic regime, relatively lower $\dot\varepsilon_s$ either in the plug zone or the localized shear band show that the temporal deformation of particles is constrained by the cohesion due to the stronger energy dissipation.

A flow depth $h$ can be extracted as the distance from the flow surface (red dashed lines) to the interface (solid lines) as illustrated in Fig. \ref{fig:strrcontour_diagram}, case $We$ = 35. The flow depth ($h$) is plotted along the $x$ axis shown in Fig. \ref{fig:flowdepth_weber_strainrate} (a). The concave shape of $h$ depicts the inhomogeneity of the flow region observed in Fig. \ref{fig:strrcontour_diagram}, which is an intrinsic feature of rotating drums \cite{orpe2001scaling,gdr2004dense}. The maximum flow depth $h_\mathrm{max}$, as a meso-scale index, is plotted against $We$ in Fig. \ref{fig:strrcontour_diagram} (b). Greater $h_\mathrm{max}$ is observed when the Weber number is small, indicating the viscoplastic regime (cohesion-dominated regime), and it is insensitive to inertial effects, \textit{i.e.}, $h_\mathrm{max}$ remains similar and $\omega$ has little effect. As $We$ increases, the flow then gradually enters the collisional regime (inertia-dominated regime), where $h_\mathrm{max}$ becomes less relevant to the cohesion force. 

\begin{figure}
\centering
\includegraphics[width=0.5\textwidth,height=\textheight,keepaspectratio]{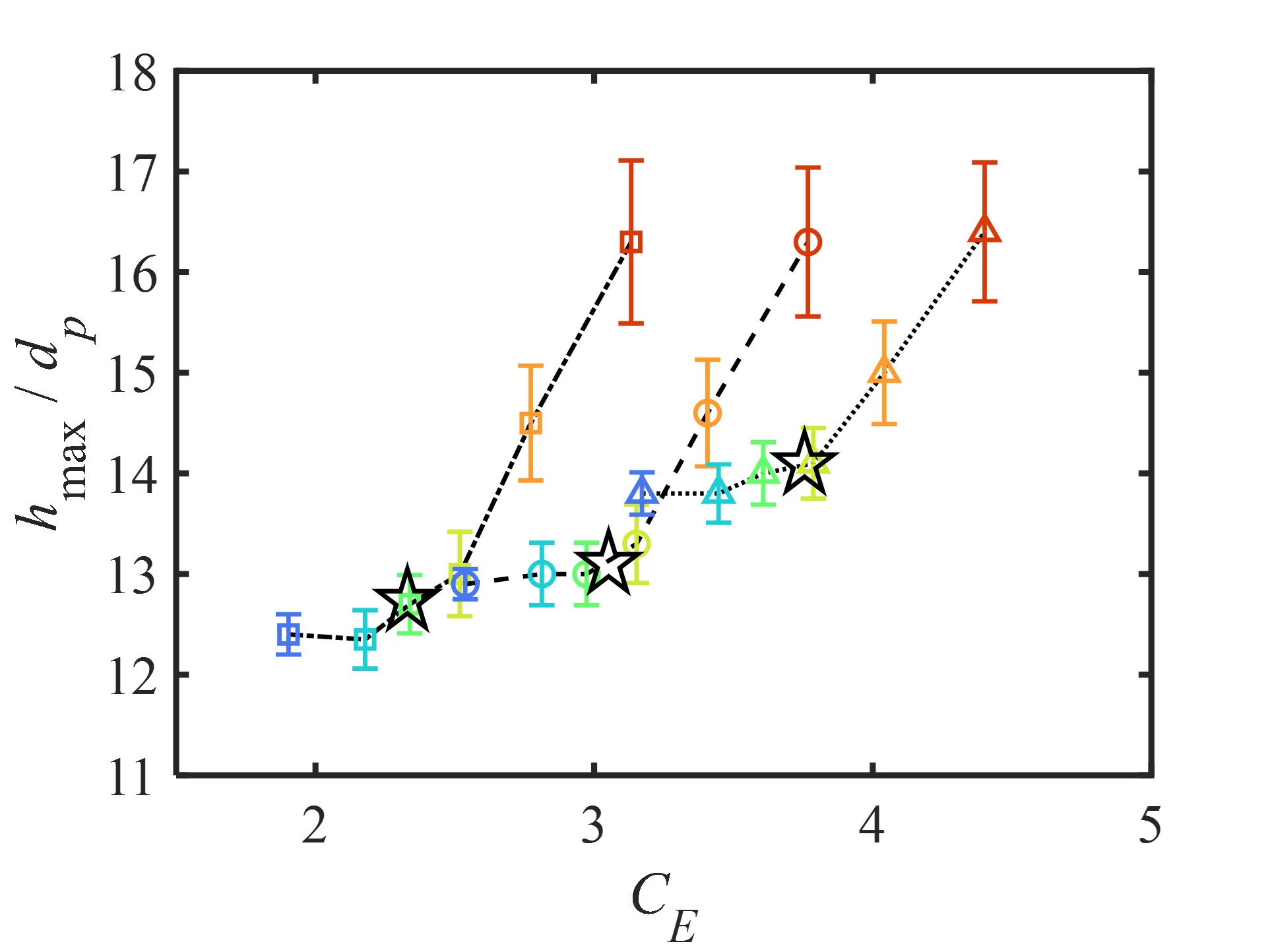}
\caption{A plot of normalized $h_\mathrm{max}$ as a function of $C_E$. Stars mark critical mode and flow regime transition points for three $\omega$ cases with $We\approx400$. The corresponding approximation of $We$ can be found in Fig. \ref{fig:strrcontour_diagram}. Colour transitions from blue to red indicate an increase in cohesion, which can be referred to Fig. \ref{fig:surfaceprofile_DAOR} (b). Dashed, dot-dashed, and dotted lines are the guide of the eye of cases $\omega$ = 15, 20 and 25 rpm respectively.}
\label{fig:flowdepth_CE}
\end{figure}

Previously we have shown that the dimensionless number $C_E$ proposed in Section \ref{subsec:anglerepose} captures the interplay effect between gravity, cohesion, and inertia, and it is effective in describing the dynamic angle of repose. Here, we plot $h_\mathrm{max}$ against $C_E$ in Fig. \ref{fig:flowdepth_CE}. It can be seen that for different rotation rate, as cohesion increases, $h_\mathrm{max}$ in all cases experience a mild rise as $C_E$ increases before a sharp climb. The threshold of Weber number $We\approx400$ discussed previously again shows good prediction in the regime transition (black stars in Fig. \ref{fig:flowdepth_CE}).

To understand the combined effects on $h_\mathrm{max}$ described by $C_E$, again, we discuss gravity, inertia and cohesion effects separately. Since the increased capillary force increases the cohesion, the solid region can possess higher strength and reach a higher peak, meaning higher gravitational potential is accumulated. Thus, when the solid region is sheared to fail and flow down, more energy needs to be dissipated compared to less cohesive cases. In Fig. \ref{fig:strrcontour_diagram}, flow region with smaller $\dot\varepsilon_s$ due to the dominance of capillary force implies a less intensive deformation, meaning the increased inter-particle bonds can enhance energy dissipation in the flow region \cite{kovalcinova2018energy}. Therefore, $C_E$ is effective in describing the role that cohesion plays in increasing the potential energy (through maintaining the solid structure) in the solid region\cite{tegzes2002avalanche} and a competing role in enhancing the energy dissipation in the flow region. In summary, a flow regime transition can be captured using the new proposed dimensionless number $C_E$. The flow regime transition, \textit{i.e.}, from collision to the viscoplastic regime, is consistent with the flow mode transition, \textit{i.e.}, from cascading to rolling mode, when cohesion increases.

\begin{figure*}
\centering
\includegraphics[width=0.8\textwidth,height=\textheight,keepaspectratio]{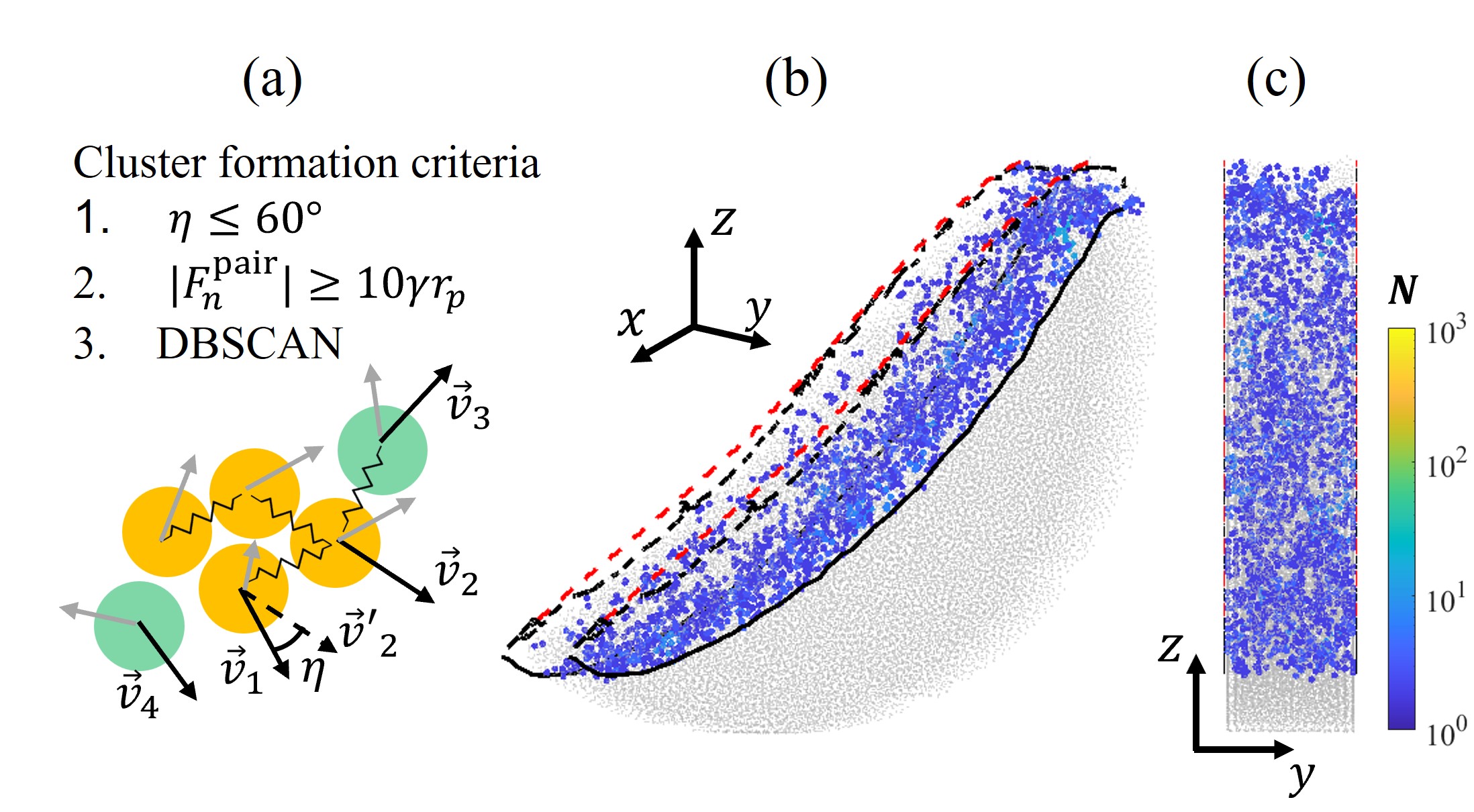}
\caption{Cluster formation criteria and an illustration example of clusters formed in granular flow. (a) The cluster formation criteria and illustrations. $\eta$ is the angle between the vector of fluctuation velocity (black arrows) of each pair of neighbours. Grey arrows depict the true velocity of each particle. The spring between particles represents the normal contact force $F_n^\mathrm{pair}$. The DBASCAN algorithm is adopted using MATLAB toolbox, which was developed by \citet{ester1996density}. Yellow circles indicate the formed clusters, and the greens are single particles. (b) and (c) show two 3D perspectives of one example of cluster formation in wet granular flow. Red circles circulate four cluster locations. The red dashed line, black dashed line, and black solid line are flow surface and $\dot\varepsilon$ boundaries the same as the case in Fig. \ref{fig:strrcontour_diagram} with $\omega$ = 15~rpm and $\gamma$ = 0.037 N/m.}
\label{fig:cluster_formation_criteria}
\end{figure*}

\begin{figure*}
\centering
\includegraphics[width=0.8\textwidth,height=\textheight,keepaspectratio]{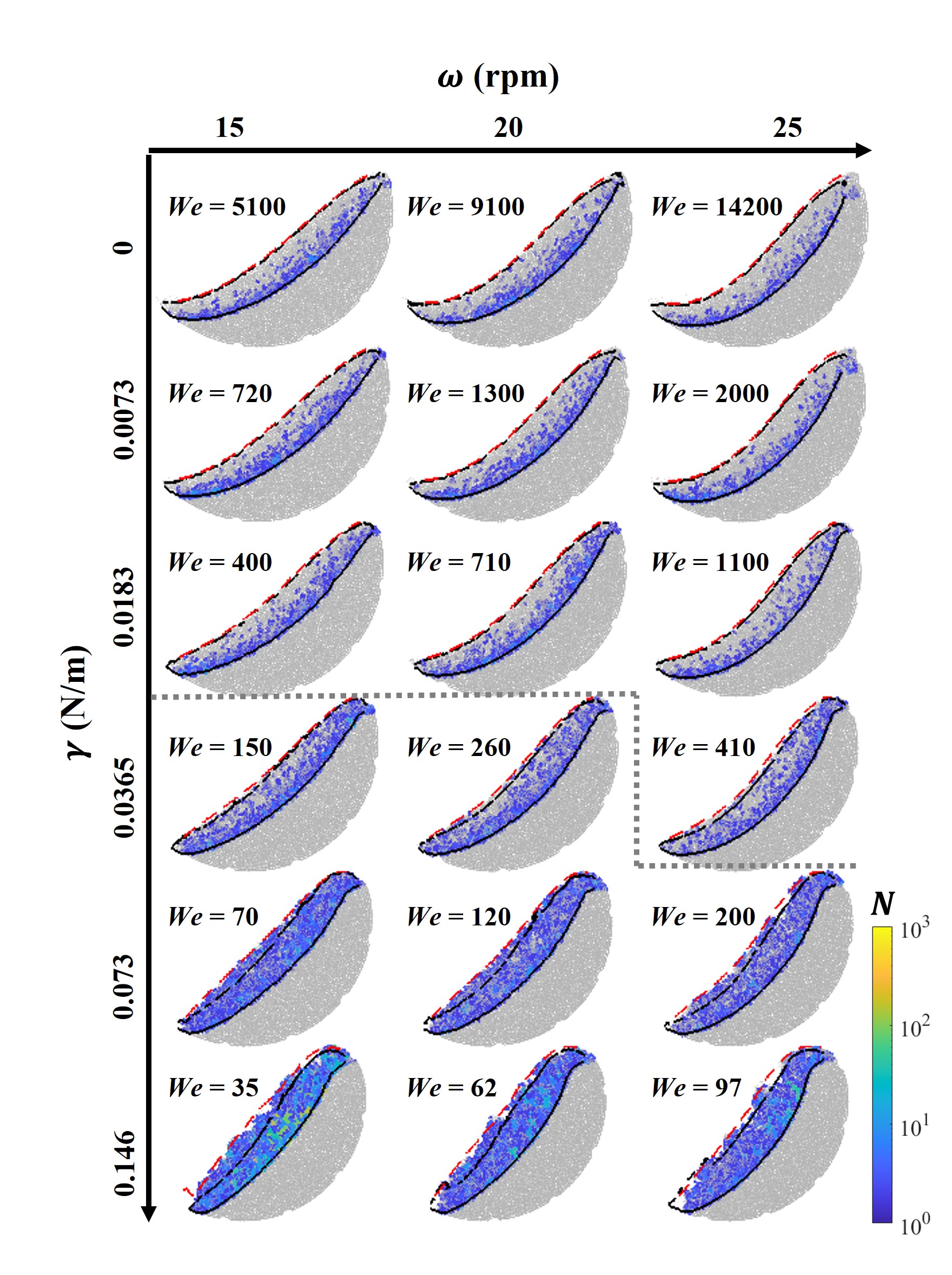}
\caption{Phase diagrams of cluster formation with varying cohesion and rotation rate. The colour scale from blue to red indicates an increase in cluster size $N$.}
\label{fig:cluster_diagram}
\end{figure*}

\subsection{Cluster formation}\label{subsection:cluster}
The discrete nature of granular flow makes it complex and challenging to characterise \cite{andreotti2013granular,murray2012continuum}. Clusters can form in granular flows, which can make the energy dissipation in the flow region more complex \cite{rognon2008dense, cao2018structural}. Here, we first introduce our cluster formation criteria modified based on previous research, followed by discussions on cluster spatial information, cluster size distribution and the relationship with flow mechanisms.

Based on the concept of velocity fluctuation, inter-particle contact and inter-particle displacement, several methods have been used to characterise cluster formation \cite{bonamy2002multiscale,campbell2011clusters,fullmer2017clustering, zou2022microscopic}. Given that cohesion is essential regarding inter-particle constraints, we adopt particle-level indices extracted from velocity fluctuation and inter-particle contact force to investigate cluster formation in the cohesive granular flow (see illustration in Fig. \ref{fig:cluster_formation_criteria} (a)). By taking these two factors into consideration, we aim to extract particles that possess relatively higher normal interaction forces and similar moving directions, which guarantee a strong inter-particle bond and collective motion. To implement the method, normal contact forces that meet the criteria $|F_n^\mathrm{pair}|\ge 10\gamma r_p$ are classified as potential cluster participants. A large coefficient of $\gamma r_p$ can lead to most particles belonging to one cluster, while a smaller one can result in no clusters being identified. The coefficient $10$ is used to capture the variation of cluster size under different cohesion cases. To describe the spatial kinetic correlation among particles, the local velocity fluctuation is calculated for each particle as $\Vec{v}_f = \Vec{v}_i - {v}_m$, where $i$ indicates the target particle and $v_m$ is the corresponding local mean velocity. The local mean velocity is averaged by taking all particles, through accumulating more than a hundred-time steps, that locate in a sampling mesh grid into consideration. An angle $\eta$ between each pair of $\Vec{v}_{f,i}$ and $\Vec{v}_{f,j}$ is calculated, and a critical angle $\eta_\mathrm{cri} \le 60^{\circ}$ is set to filter out non-correlated particles\cite{bonamy2002experimental} (see Fig. \ref{fig:cluster_formation_criteria} (a)). The DBSCAN algorithm is applied to construct clusters based on potential cluster participants screened by criteria $\eta$ and $F_n^\mathrm{pair}$. The DBSCAN algorithm sequentially searches for neighbours of a target particle to form clusters until no outside-cluster neighbouring participants can be found, which can avoid relatively weak connections (single particle connection) among clusters \cite{ester1996density,zou2022microscopic}. 

A cluster formation illustration is shown in Fig. \ref{fig:cluster_formation_criteria} (a), where cluster particles are coloured yellow, and single particles are coloured green. A 3D view and a front view of an example where clusters forms in a simulation case ($\omega = 15~\mathrm{rpm}$ and $\gamma = 0.0365~\mathrm{N/m}$) are shown in Fig. \ref{fig:cluster_formation_criteria} (b) and (c), where clusters are formed in the flow region. The cluster formation under different cohesion and rotation rate is shown in Fig. \ref{fig:cluster_diagram}, which illustrates that clusters occupy the region near the flow-solid interface in relatively fewer cohesive cases while spreading across the whole flow region in relatively strong cohesive cases. A reason for this can be that relatively weak cohesive contacts tend to develop near the surface region where no overburden pressure is applied. The weak contacts are easier to break, so clusters are rarely formed in that region. In strongly cohesive cases, it can be found that relatively large clusters appear mostly in the shear band (underneath the plug zone), which supports the view that the cluster formation in the shear band of cohesive granular materials is vital to resist shear\cite{rognon2008dense}. The appearance of clusters near the flow surface indicates the reason for a repetitive fluctuating surface profile shown in the corresponding plots of cases $We<400$ in Fig. \ref{fig:cluster_diagram} and in work by \citet{tegzes2002avalanche}. Under this condition, an increase in rotation rate is found to have little effect on the cluster-occupied area and cluster sizes due to the cohesion-dominant mechanism.

\begin{figure*}
\centering
\includegraphics[width=0.9\textwidth,height=\textheight,keepaspectratio]{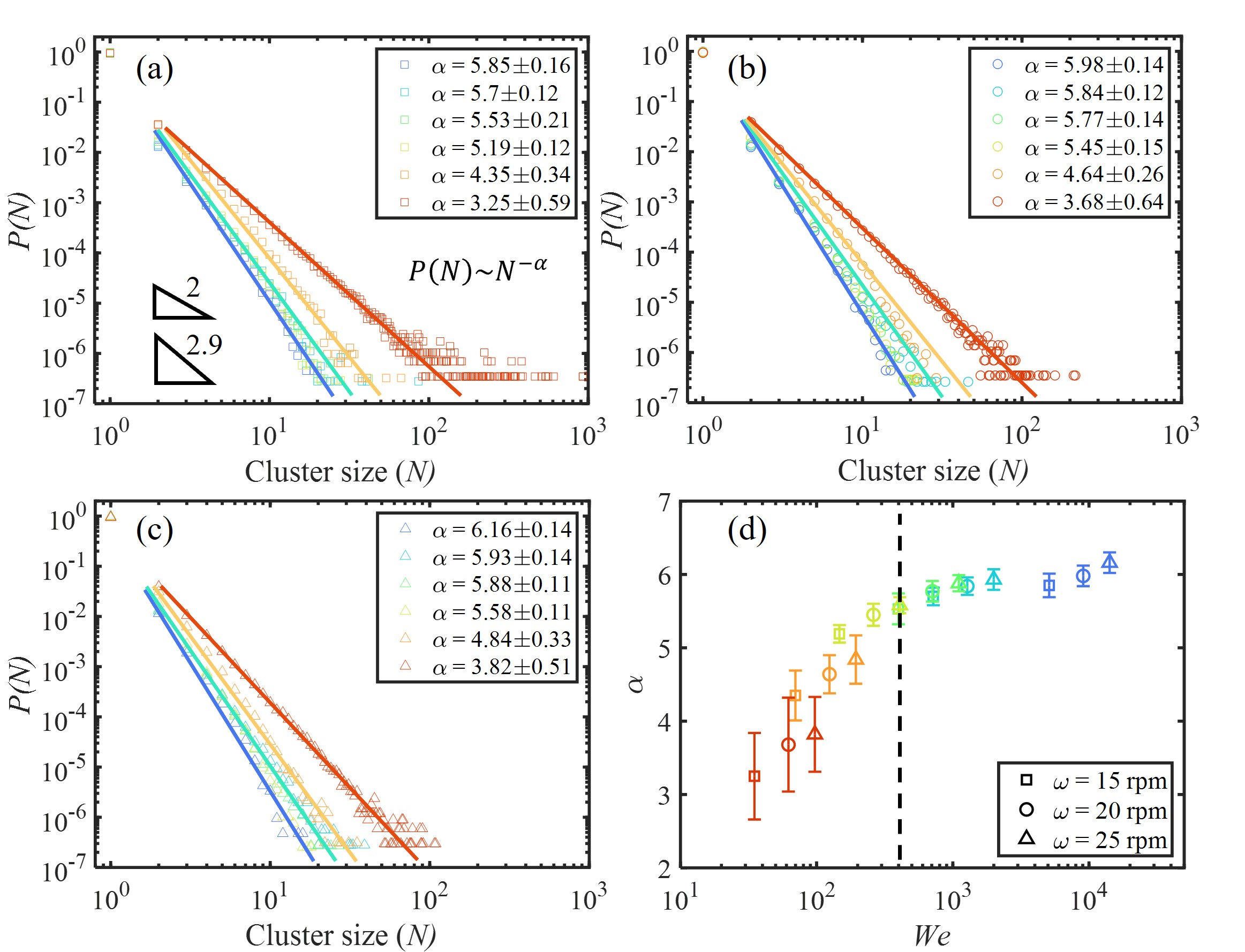}
\caption{Cluster size distribution. (a)-(c) The probability distribution ($P$) of cluster size ($N$). $\square$, {\Large $\circ$}, and $\triangle$ represent $\omega$  = 15, 20 and 25~rpm respectively. The colour transition from blue to red indicates an increase in cohesion, which can be referred to Fig. \ref{fig:surfaceprofile_DAOR} (b). Solid lines are the guide of the eye for the power law relation shown in (a). Triangles in (a) indicate $\alpha$ slopes of the 2D ($\alpha = 2\pm0.1$) and quasi-2D ($\alpha = 2.9\pm0.1$) cases for mono-sized spherical particles under dry conditions in the work by \citet{bonamy2002multiscale}. (d) The power index $\alpha$ is plotted as the function of $We$, where the colour transition is consistent with (a)-(c).}
\label{fig:cluster_distribution_alphaVSWe}
\end{figure*}

Corresponding to the phase diagram of cluster formation in Fig. \ref{fig:cluster_diagram}, the cluster size distribution of each case is plotted in Fig. \ref{fig:cluster_distribution_alphaVSWe} (a)-(c). A power law decay of the probability of cluster size $P(N) \propto N^{-\alpha}$ is found to be consistent with similar 2D and quasi-2D cases\cite{bonamy2002multiscale} with $\omega = 8~\mathrm{rpm}$, where a smaller value of $\alpha$ indicates a greater chance of the formation of large clusters. For the cohesionless cases, the results by \citet{bonamy2002multiscale} together with our results show that a higher level of boundary constraints (from 3D to 2D) and a lower rotation rate (less kinetic energy) both make the flow tend to form larger clusters. This demonstrates that the velocity fluctuation can be constrained by more strict boundary confinement and lower inertia effects \cite{pouliquen2004velocity,fullmer2017clustering}. As the cohesion rises, according to Fig. \ref{fig:cluster_distribution_alphaVSWe} (a)-(c), $\alpha$ decreases, implying a general increase in the size of formed clusters. Finally, $\alpha$ is plotted against $We$ to describe the competing effect between cohesion energy and kinetic energy when observing cluster formation. Same as Fig. \ref{fig:flowdepth_weber_strainrate} (b) and Fig. \ref{fig:flowdepth_CE}, the result shows that $We\approx400$ is a critical value above which the effects from cohesion on cluster size becomes negligible.

\begin{figure}
\centering
\includegraphics[width=0.5\textwidth,height=\textheight,keepaspectratio]{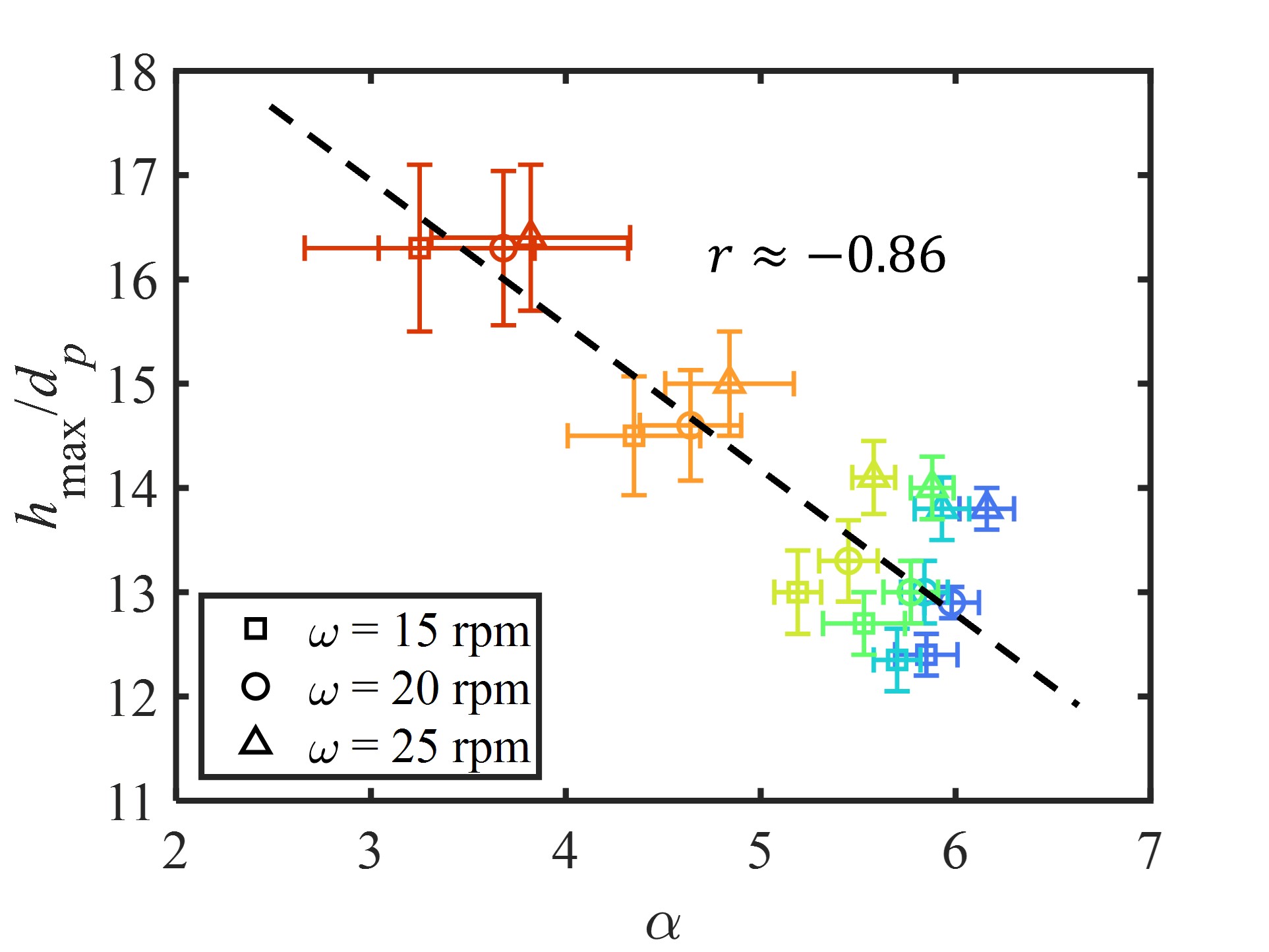}
\caption{The max flow depth $h_\mathrm{max}$ versus cluster size distribution power index $\alpha$ shown in Fig. \ref{fig:cluster_distribution_alphaVSWe}. The dashed line is a line of best fit with correlation coefficient $r$.}
\label{fig:Dmax_alpha}
\end{figure}
 
To clarify the relationship between cluster formation and granular flow regime, the maximum flow depth $h_\mathrm{max}$ is plotted against $\alpha$ shown in Fig. \ref{fig:Dmax_alpha}. The Pearson coefficient $r \approx 0.86$ implies that the cluster size and flow depth are in a considerable positive correlation relationship. Since the clusters can possess a relatively irregular topology compared to single-sphere particles, higher resistance of inter-cluster contact (higher inter-cluster friction and rolling resistance) can be raised to hinder the relative inter-cluster movement \cite{rognon2008dense,campbell2011clusters}. Moreover, \citet{cao2018structural} also mentioned that the inter-cluster and intra-cluster plastic rearrangement can play key roles in dissipating energy in granular flows. In summary, an increase in cohesion is found to induce the granular flow to form relatively larger clusters which can be the key regime affecting the granular flow.

\section{Conclusions}
In this study, partially saturated granular flow in a rotating drum is studied experimentally and numerically. We first demonstrate that the experimentally observed rolling and cascading modes, controlled by the interparticle cohesion and drum rotation rate, are successfully captured in the numerical simulation using the Discrete Element Method. 

With the combined effects of cohesion, inertia, and gravity, a dimensionless parameter, $C_E$, is proposed to quantitatively describe the cohesive granular flow. For the first time, a collapse of data on the dynamic angle of repose under different rotation rate and cohesive forces are observed. This implies the capability and universality of the proposed $C_E$ in capturing the interplay among key mechanisms for cohesive granular flows. 

At the mesoscale and particle level, the depth of flow region and the statistics of cluster formations in cohesive granular flow are examined. The transition from an inertia-dominated regime to a cohesion-dominated regime is highlighted. We further explored the correlation between the flow depth and cluster formation (with a correlation coefficient > 0.8), showing the effects of the cohesion manifested by the cluster formation.

Our results and analyses shed light on the multiscale connection across the particle-level mechanism, \textit{i.e.}, grain agglomeration,  expansion of flow region (meso), and macroscopic observations characterised by the dynamic angle of repose. 
This study provides a new perspective on cohesive granular flows via revealing the dominating lower-scale features. This could benefit the development of various engineering applications, such as the food industry, mineral handling, and slope analysis under unsaturated conditions.

\bibliography{aipsamp.bib}

\end{document}